\documentclass[10pt]{article} % For LaTeX2e

\usepackage[numbers,sort&compress]{natbib}
\usepackage[preprint]{neurips_2025}

\usepackage[utf8]{inputenc} % allow utf-8 input
\usepackage[T1]{fontenc}    % use 8-bit T1 fonts
\usepackage{hyperref}       % hyperlinks
\usepackage{url}            % simple URL typesetting
\usepackage{booktabs}       % professional-quality tables
\usepackage{amsfonts}       % blackboard math symbols
\usepackage{nicefrac}       % compact symbols for 1/2, etc.
\usepackage{microtype}      % microtypography
\usepackage{xcolor}         % colors

\usepackage{amsmath,amsfonts,bm}

\def\eqref#1{equation~\ref{#1}}
\def\1{\bm{1}}

\DeclareMathAlphabet{\mathsfit}{\encodingdefault}{\sfdefault}{m}{sl}
\SetMathAlphabet{\mathsfit}{bold}{\encodingdefault}{\sfdefault}{bx}{n}

\usepackage{amsthm}

\newtheorem*{proposition*}{Proposition}

\newtheorem*{claim*}{Claim}
\theoremstyle{definition}

\newtheorem*{definition*}{Definition}

\usepackage{graphicx}       % Required for inserting images
\usepackage{subcaption}
\usepackage{bbm}
\usepackage{amssymb}% http://ctan.org/pkg/amssymb
\usepackage{bm} % bold math
\usepackage{pifont}% http://ctan.org/pkg/pifont
\usepackage{tikz}
\usepackage{colortbl}

\usepackage{booktabs}       % professional-quality tables
\usepackage{multirow}
\usepackage{outlines}

\usepackage{pgfplots}
\usepgfplotslibrary{groupplots}
\usepackage{filecontents}
\pgfplotsset{compat=newest}
\usetikzlibrary{pgfplots.fillbetween}% 
\usetikzlibrary{patterns, patterns.meta}
\pgfplotsset{compat=newest}
\usetikzlibrary{backgrounds,arrows.meta,hobby}
\usetikzlibrary{arrows, decorations.markings}
\usepgfplotslibrary{colorbrewer}

\usepackage{tikz-3dplot}
\colorlet{estColor}{green!40!black}
\colorlet{errColor}{red}
\colorlet{ubColor}{blue}
\colorlet{daubColor}{green!70!black}
\colorlet{legColor}{blue!80!black}
\colorlet{fouColor}{yellow!70!red}
\tikzset{>=latex} % for LaTeX arrow head
\tikzstyle{vector}=[-stealth,thick,line cap=round]
\usepackage{pgfplotstable}
\usetikzlibrary{decorations.pathreplacing,calligraphy}
\usetikzlibrary{fit, calc, positioning}

\def\teasertpath{Figures/NLA-teasers/Teaser_translated_5.csv}

\title{WaLRUS: Wavelets for Long-range Representation Using SSMs}

\author{%
  Hossein Babaei\\
  Department of Electrical and Computer Engineering\\
  Rice University\\
  \texttt{hb26@rice.edu} \\
   \And
   Mel White \\
   Department of Electrical and Computer Engineering\\
   Rice University\\
   \texttt{mel.white@rice.edu} \\
   \And
   Sina Alemohammad \\
   Department of Electrical and Computer Engineering\\
   Rice University\\
   \texttt{sa86@rice.edu} \\
   \And
   Richard G. Baraniuk \\
   Department of Electrical and Computer Engineering\\
   Rice University\\
   \texttt{richb@rice.edu} \\
   }

\usepackage{amsthm}
\theoremstyle{plain}

\begin{document}

\maketitle

\begin{abstract}
State-Space Models (SSMs) have proven to be powerful tools for modeling long-range dependencies in sequential data. 
While the recent method known as HiPPO has demonstrated strong performance, and formed the basis for machine learning models S4 and Mamba, it remains limited by its reliance on closed-form solutions for a few specific, well-behaved bases. 
The SaFARi framework generalized this approach, enabling the construction of SSMs from arbitrary frames, including non-orthogonal and redundant ones, thus allowing an infinite diversity of possible ``species'' within the SSM family. 
In this paper, we introduce WaLRUS (Wavelets for Long-range Representation Using SSMs), a new implementation of SaFARi built from Daubechies wavelets. 
%We instantiate two variants, scaled-WaLRUS and translated-WaLRUS, and show that the multiresolution and localized nature offers significant advantages in representing non-smooth and transient signals. 
% We instantiate two variants, scaled-WaLRUS and translated-WaLRUS, and show that the advantages of wavelets in representing non-smooth and transient signals are preserved when implemented by SSM. 
We compare WaLRUS to HiPPO-based models and demonstrate improved accuracy and more efficient implementations for online function approximation tasks.\footnote{To facilitate reproducibility, we provide code and supplementary material at the following anonymous repository:  
\url{https://osf.io/7kjcx/?view_only=5dc38b9776624deb9d1c0d8f88108658}. }
\end{abstract}

\section{Introduction}
Sequential data is foundational to many machine learning tasks, including natural language processing, speech recognition, and video understanding \cite{eventcam2024, alemohammad2019wearing,S4ND_2022}. 
These applications require models that can effectively process and retain information over long time horizons. 
A central challenge in this setting is the efficient representation of long-range dependencies in a way that preserves essential features of the input signal for downstream tasks, while remaining computationally tractable during both training and inference \cite{hochreiter1997long}.

Recurrent neural networks (RNNs) are traditional choices for modeling sequential data, but struggle with long-term dependencies due to vanishing or exploding gradients during backpropagation through time \cite{ElmanRNN, hochreiter1997long, schuster1997bidirectional}. 
While gated variants like LSTMs \cite{graves_lstm_2005} and GRUs \cite{gru} mitigate some issues, they require significant tuning and lack compatibility with parallel processing, hindering scalability.

%While deep learning models such as recurrent neural networks (RNNs) have been widely adopted for modeling sequential data, they are fundamentally limited in their ability to capture long-range dependencies \cite{ElmanRNN, hochreiter1997long, schuster1997bidirectional}.
%Backpropagation through time propagates gradients across repeated weight matrices, leading to vanishing or exploding gradients. These phenomena severely restrict the ability of RNNs to store long-term context, impairing both training stability and generalization. Gated variants like LSTMs \cite{graves_lstm_2005} and GRUs \cite{gru} alleviate some of these issues, yet they still demand extensive tuning, lack generalization across variable-length sequences, and inherently rely on sequential computation, which limits their compatibility with parallel processing on modern hardware.

State-space models (SSMs) offer a linear and principled framework for encoding temporal information, and have re-emerged as a powerful alternative for online representation of sequential data \cite{gu2020hippo,gu2022s4,gu2023mamba,gu2022train,diag_effective_2024,diag_init_2022,smith2023s5,liquid_s4_2023}. 
By design, they enable the online computation of compressive representations that summarize the entire input history using a fixed-size state vector, ensuring a constant memory footprint regardless of sequence length.
A major breakthrough came with HiPPO (High-order Polynomial Projection Operators), which reformulates online representation as a function approximation problem using orthogonal polynomial bases \cite{gu2020hippo}. 
%HiPPO enables compact, real-time updates of signal representations, and forms the foundation of several state-of-the-art architectures for long-range modeling, including S4 and Mamba \cite{gu2022s4,gu2023mamba}. These models have shown remarkable performance in language, vision, and control applications due to their ability to combine memory efficiency with theoretical rigor.
This approach underpins state-of-the-art models like S4 and Mamba, enabling compact representations for long-range dependencies \cite{gu2022s4,gu2023mamba}.

However, existing SSMs primarily rely on Legendre and Fourier bases, which, although effective for smooth or periodic signals, struggle with non-stationary and localized features\cite{gu2020hippo,gu2022s4}. 
These challenges are especially evident in domains such as audio, geophysics, and biomedical signal processing, where rapid transitions and sparse structure are common.
%Despite their success, existing SSM formulations are constrained by their reliance on a small number of well-behaved bases, primarily Legendre and Fourier, which, while effective for smooth or periodic functions, struggle with signals that contain localized features, singularities, or non-stationarities \cite{gu2020hippo,gu2022s4}. These challenges are especially evident in domains such as audio, geophysics, and biomedical signal processing, where rapid transitions and sparse structure are common.

To address this limitation, the SaFARi framework (State-Space Models for Frame-Agnostic Representation) extends HiPPO to arbitrary frames, including non-orthogonal and redundant bases \cite{diag_effective_2024,diag_init_2022, babaei2025safari}. 
This generalization enables SSM construction from any frame via numerical solutions of first-order linear differential equations, preserving HiPPO's memory efficiency and update capabilities without closed-form restrictions.

%To address this limitation, we adopt and extend the SaFARi framework (State-Space Models for Frame-Agnostic Representation), which generalizes HiPPO by enabling the use of arbitrary frames or bases, including non-orthogonal and redundant ones \cite{diag_effective_2024,diag_init_2022}. SaFARi \cite{babaei2025safari} formulates online function approximation as the numerical solution of a first-order linear differential equation, deriving the update matrices $A(t)$ and $B(t)$ directly from any chosen frame. This numerical approach decouples SSM construction from the need for closed-form expressions, significantly broadening the design space for state-space models while preserving the desirable properties of HiPPO, such as constant memory, efficient updates, and compatibility with downstream neural networks.

 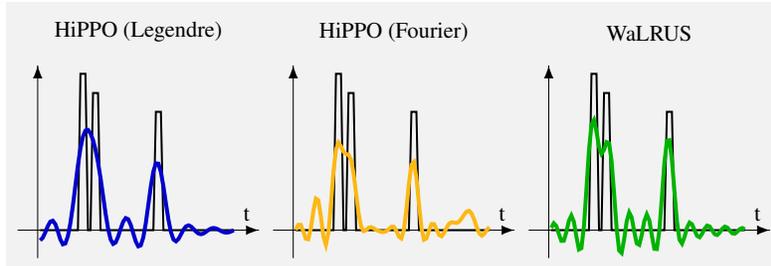
\begin{figure}[t!]
     \centering
     \resizebox{0.75\textwidth}{!}{%}
     \begin{tikzpicture}[background rectangle/.style={fill=gray!10},
    show background rectangle]
\centering
\makeatletter
\tikzset{ nomorepostaction/.code=\makeatletter\let\tikz@postactions\pgfutil@empty, % From https://tex.stackexchange.com/questions/3184/applying-a-postaction-to-every-path-in-tikz/5354#5354
    my axis/.style={
        postaction={
            decoration={
                markings,
                mark=at position 1 with {
                    \arrow[thick]{latex}
                }
            },
            decorate,
            nomorepostaction
        },
        thin,
        -, % switch off other arrow tips
        every path/.append style=my axis % this is necessary so it works both with "axis lines=left" and "axis lines=middle"
    }
}
\makeatother

\pgfplotsset{/pgfplots/group/every plot/.append style = {
    xmin=0,
    xmax=80, % axis limits
    ymin=-.004, % have to fix y limits,
    ymax=.05, % otherwise x axes don't line up
    restrict x to domain=1:78, % so that the data doesn't go all the way to the end of the axis
    axis lines=center,
    anchor=origin,
    grid,xtick=\empty,
    ytick=\empty,
    axis line style={my axis},
    thick,
    enlargelimits=true
}};

\begin{groupplot}[group style = {group size = 3 by 1, horizontal sep = 2mm},width=5cm]%, height = 5.0cm]
  
    \nextgroupplot[
        title ={\small HiPPO (Legendre)},
        xlabel={\small t},
        ]
        \addplot[]
        table [
            x index=0, 
            y index=1, 
            col sep=comma] 
            {\teasertpath};
        \addplot[legColor,ultra thick
        ]
        table [
            x index=0,
            y index=3,
            col sep=comma]
            {\teasertpath};
        ]
        
    \nextgroupplot[
        title ={\small HiPPO (Fourier)},
        xlabel={\small t},
    ]
        
        \addplot[]
        table [
            x index=0, 
            y index=1, 
            col sep=comma] {\teasertpath};
        \addplot[fouColor,ultra thick
        ]
        table [
            x index=0,
            y index=4,
            col sep=comma]
            {\teasertpath};
        ]
        
\nextgroupplot[
        title ={\small WaLRUS },
        xlabel={\small t},]

        \addplot[]
        table [
            x index=0, 
            y index=1, 
            col sep=comma] {\teasertpath};
        
        \addplot[daubColor, ultra thick
        ]
        table [
            x index=0,
            y index=2,
            col sep=comma]
            {\teasertpath};
        ]

\end{groupplot}

\end{tikzpicture}
     }
     \caption{An input signal comprising three random spikes is sequentially processed by SSMs and reconstructed after observing the entire input.
     Only the wavelet-based SSM constructed using WaLRUS can clearly distinguish adjacent spikes.
     \label{fig:teaser}
 }
\end{figure}

In this paper, we leverage SaFARi with wavelet frames to introduce WaLRUS (Wavelets for Long-range Representation Using SSMs). 
We propose two variants: scaled-WaLRUS and translated-WaLRUS, designed for capturing non-smooth and localized features through compactly supported, multi-resolution wavelet decompositions \cite{daubechies1992ten}. 
These properties allow WaLRUS to retain fine-grained signal details typically lost in polynomial-based models.

%In this paper, we instantiate SaFARi using wavelet frames, yielding a new class of models that we call WaLRUS (Wavelets for Long-range Representation Using SSMs). Specifically, we introduce two variants: scaled-WaLRUS for the scaled uniform measure, and translated-WaLRUS for the translated measure. Wavelets are particularly well-suited for representing non-smooth or localized structures due to their compact support and multi-resolution decomposition \cite{daubechies1992ten}. These characteristics allow Walrus models to retain fine-grained features in the signal that polynomial-based models typically lose.

As a canonical example, we derive WaLRUS using Daubechies wavelets, and provide a rigorous comparative analysis of WaLRUS and existing HiPPO variants (see Fig.~\ref{fig:teaser}). 
% We provide a rigorous comparative analysis of WaLRUS and existing HiPPO variants. 
%Empirical results demonstrate that scaled-WaLRUS and translated-WaLRUS consistently outperform Legendre and Fourier-based models in reconstruction accuracy, especially on signals with sharp transients. 
Empirical results demonstrate that the wavelet-based WaLRUS model consistently outperforms Legendre and Fourier-based HiPPO models in reconstruction accuracy, especially on signals with sharp transients. 
Furthermore, WaLRUS enjoys diagonalizability, which is the key enabler of efficient convolution-based implementations and parallel computation \cite{diag_effective_2024,diag_init_2022}.

% As a canonical example, we derive WaLRUS using Daubechies wavelets, and demonstrate improved reconstruction of spiky, transient signals compared to HiPPO models (see Fig.~\ref{fig:teaser}). 
These results highlight the practical advantages of WaLRUS models, particularly in scenarios where signal structure varies across time and scale.
% In summary, this paper introduces WaLRUS, the first integration of wavelet frames into the SSM paradigm via SaFARi. 
% We show that this generalization not only enhances the expressiveness of online representations but also yields tangible benefits in accuracy and computational efficiency. 
By bridging multiscale signal analysis and online function approximation, WaLRUS opens new directions for modeling complex temporal phenomena across disciplines.

%This paper is organized as follows. 
%In Section \ref{sec:background}, we review the HiPPO framework and its limitations, motivating the need for a generalized approach. 
%Section \ref{sec:math_prelim} provides the required mathematical preliminaries. 
%Section \ref{sec:SAFARI} introduces our frame-agnostic method for SSM construction, and then Section \ref{sec:SAFARI-Daub} follows this method to derive the wavelet-based SaFARi model. 
%Section \ref{sec:implementation} addresses the implementation considerations and strategies for SaFARi, including the approximation of its infinite-dimensional representation in finite dimensions, and provides a rigorous theoretical analysis of the associated errors. 
%Section \ref{sec:experiments} presents experimental results, comparing SaFARi to existing methods and analyzing its performance. 
%Finally, Section \ref{sec:conclusions} discusses the broader implications of our work and outlines directions for future research.

\section{Background}
Recent advances in machine learning, computer vision, and large language models have pushed the frontier of learning from long sequences of data. These applications demand models that can (1) generate compact representations of input streams, (2) preserve long-range dependencies, and (3) support efficient online updates.

Classical linear methods, such as the Fourier transform, offer compact representations in the frequency domain \cite{oppenheim1999discrete,abbate2012wavelets,box2015time,proakis2001digital,prandoni2008signal}. However, they are ill-suited for online processing: each new input requires recomputing the entire representation, making them inefficient for streaming data and limited in their memory horizon. Nonlinear models like recurrent neural networks (RNNs) and their gated variants (LSTMs, GRUs) have been more successful in sequence modeling, but they face well-known issues such as vanishing/exploding gradients and limited parallelization  \cite{ElmanRNN,hochreiter1997long,gru,schuster1997bidirectional}. 
Moreover, their representations are task-specific, and not easily repurposed across different settings.

To resolve these issues, the HiPPO framework \cite{gu2020hippo} casts online function approximation as a continuous projection of the input $u(t)$ onto a linear combination of the given basis functions $\mathcal{G}$. %over the interval $[0, T]$. 
At every time $T$, it solves $\min_{g^{(T)} \in \mathcal{G}} \| u_T - g^{(T)}(t) \|_\mu$ , producing a compressed state vector $\vec{c}(T)$ that satisfies the update rule:
\begin{equation}
   \frac{d}{dT} \vec{c}(T) = -A_{(T)} \vec{c}(T) + B_{(T)} u(T). 
   \label{Time_Varying_SSM}
\end{equation}

Here, $A_{(T)}$ and $B_{(T)}$ are derived based on the choice of polynomial basis and measure $\mu(t)$, which defines how recent history is weighted. Two commonly used measures are:
\begin{equation}
    \mu_{\textit{tr}}(t)= \frac{1}{\theta}\mathbbm{1}_{t\in[T-\theta,T]}, \quad 
    \mu_{\textit{sc}}(t)= \frac{1}{T}\mathbbm{1}_{t\in[0,T]}.  \label{Measures}
\end{equation}
The translated measure $\mu_{\textit{tr}}$ emphasizes recent history within a sliding window of length $\theta$, while the scaled measure $\mu_{\textit{sc}}$ compresses the entire input history into a fixed-length representation.

Despite its strengths, HiPPO is restricted to only a few bases (e.g., Legendre, Fourier), and deriving $A(t)$ and $B(t)$ in closed form is only tractable for specific basis-measure combinations.

SaFARi addressed this limitation by generalizing online function approximation to any arbitrary frame \cite{babaei2025safari}. A frame $\Phi(t)$ is a set of elements $\{ \phi_i(t)\}$ such that one can reconstruct any input $g(t)$ by knowing the inner products $\langle g(t), \phi_i(t)\rangle$. 
For a given frame $\Phi$, its inverse $\overline{\Phi}$, and its dual $\widetilde{\Phi}$, the scaled-SaFARi produces an SSM with the $A$ and $B$ given by:
\begin{equation}
    \frac{\partial}{\partial T} \vec{c} (T) = - \frac{1}{T} A \vec{c} (T)+ \frac{1}{T} B u(T), \quad A_{i,j}\hspace{-0.05cm}=\delta_{i,j} + \hspace{-0.1cm} \int_0^1 \hspace{-0.1cm} t' \frac{\partial}{\partial t}\overline{\phi}_{i} \bigg|_{t=t'} \hspace{-0.1cm} \widetilde{\phi}_{j} (t') dt' , \quad B_i= \overline{\phi}_i (1)
    \label{Scaling_SSM}
\end{equation}

while the translated-SaFARi produces an SSM with the $A$ and $B$ given by:
\begin{equation}
    \frac{\partial}{\partial T} \vec{c} (T) = - \frac{1}{\theta} A \vec{c} (T)+ \frac{1}{\theta} B u(T), \quad  A_{i,j}= \overline{\phi}_i (0) \widetilde{\phi}_j (0)   +\hspace{-0.1cm} \int_0^1 \hspace{-0.1cm}\frac{\partial}{\partial t} 
 \overline{{\phi}}_{i}\bigg|_{t=t'} \hspace{-0.1cm} \widetilde{\phi}_{j} (t') dt' ,\, \,  B_i= \overline{\phi}_i (1)
    \label{eq:Translate_SSM}
\end{equation}

\textbf{Incremental update of SSMs}: 
The differential equation in Eq.~\ref{Time_Varying_SSM} can be solved incrementally. Several update rules are discussed in \cite{butcher1987numerical}. Following~\cite{gu2020hippo}, we adopt the Generalized Bilinear Transform (GBT)~\citep{Zhang_GBT} given by Eq.~\ref{Update_GBT} for its superior numerical accuracy in first order SSMs. \vspace{0.1cm}
\begin{equation}
    c(t+\Delta t) = (I + \delta t \alpha A_{t+\delta t})^{-1} \left[ (I - \delta t (1-\alpha) A_t) c(t) + \delta t B(t) u(t) \right]
    \label{Update_GBT}
\end{equation}

\textbf{Diagonalization of \textbf{\textit{A}}}: 
Each GBT step involves matrix inversion and multiplication. 
If $A(t)$ has time-independent eigenvectors (e.g., $A(t) = g(t) A$), it can be diagonalized as $A(t) = V \Lambda(t) V^{-1}$, allowing a change of variables $\widetilde{c} = V^{-1}c$ and $\widetilde{B} = V^{-1} B(t)$, yielding:
\begin{equation}
    \frac{\partial}{\partial t} \widetilde{c} = -\Lambda(t) \widetilde{c} + \widetilde{B} u(t),
    \label{Diagonal_SSM}
\end{equation}
This reduces each update to elementwise operations, significantly lowering computational cost.

%\textbf{Convolution kernels and diagonalization}
%The computational complexity of sequential updates for SSMs was thoroughly analyzed in \cite{babaei2025safari}. For a sequence of length $L$, scaled-SaFARi requires $O(N^3 L)$ complexity due to the need to solve an $N$-dim linear system at each step, while translated-SaFARi has $O(N^2 L)$ complexity by reusing inverses across time steps. A key advantage emerges when the state matrix $A$ is diagonalizable, allowing both variants to reduce complexity to $O(N L)$ and leverage parallel hardware for independent scalar SSM updates, further accelerating the computation.

%However, sequential updates become prohibitively slow as the sequence length increases, particularly during the training phase when the entire sequence is available upfront. To address this, representation can be computed through convolution kernels instead of step-wise updates. By first calculating the convolution kernel and then convolving it with the input sequence, the computational process is significantly accelerated. This method particularly benefits from GPU-based parallelism, achieving $O(N \log L)$ run-time complexity when the SSM is diagonalizable. 

\subsection{Wavelet Frames}\label{sec:wavelet_frames}

% State Space Models (SSMs) provide an efficient mechanism for online function approximation by computing a representation in a fixed basis. 
% The resulting state vector can be leveraged for tasks such as reconstruction, feature extraction, and denoising, all of which benefit from compact and informative signal representations. 
Although any orthonormal basis for $L^2([0,1])$ suffices in theory to construct an SSM with SaFARi, practical performance varies significantly depending on truncation behavior and nonlinear approximation properties.

Wavelet frames offer a multiresolution analysis that captures both temporal and frequency characteristics of signals, making them particularly effective for representing non-stationary or long-range dependent data \cite{flandrin_abry}. 
Initiated by \cite{haar1909} and formalized by \cite{grossman_morlet}, wavelet theory gained prominence with Ingrid Daubechies' seminal work \cite{daubechies1988orthonormal}, which introduced compactly supported orthogonal wavelets. 
Since then, wavelets have played a central role in modern signal processing \cite{mallat_tour}.

Wavelet analysis decomposes a signal $f(t)$ into dilations and translations of a mother wavelet $\psi(t)$, enabling simultaneous localization in time and frequency. The \emph{continuous wavelet transform } is 
\[
W(a, b) = \int_{-\infty}^\infty f(t) \psi^*_{a,b}(t) \, dt,
\quad
\psi_{a,b}(t) = \frac{1}{\sqrt{a}} \psi\left(\frac{t-b}{a}\right),
\]
while the \emph{discrete wavelet transform (DWT)} uses a dyadic grid $a=2^{-j}, \, b=k$.

Unlike global bases such as Fourier or polynomials, which struggle with localized discontinuities, wavelets provide sparse representations of signals with singularities, such as jumps or spikes \cite{mallat1992singularity, daubechies1992ten}. 
Their local support yields small coefficients in smooth regions and large coefficients near singularities, enabling efficient compression and accurate reconstruction. 
These properties make wavelet frames a natural and powerful choice for time-frequency analysis in a wide range of practical applications.

\section{WaLRUS: Wavelet-based SSMs}\label{sec:SAFARI-Daub}

Daubechies wavelets \cite{daubechies1988orthonormal,daubechies1992ten} provide a particularly useful implementation of a SaFARi SSM.
While there are different types of commonly used wavelets, Daubechies wavelets are of particular interest in signal representation due to their maximal vanishing moments over compact support.

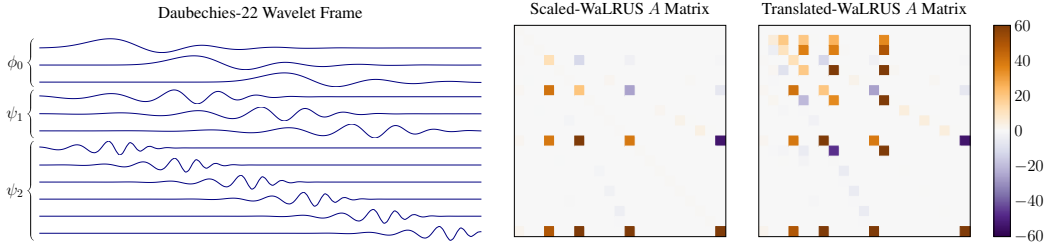
\begin{figure}[t!]
   \centering
        \resizebox{\textwidth}{!}{%}
        % \documentclass[tikz]{standalone}

% \usepackage{tikz}
% \usepackage{pgfplots}
% \usepackage{pgffor}
% \usepackage{filecontents}
% \usepackage{xcolor}
% \usetikzlibrary{pgfplots.groupplots}
% \pgfplotsset{compat=newest}
% \usepgfplotslibrary{colorbrewer}
% \usetikzlibrary{calc}
% \usetikzlibrary{decorations.pathreplacing,calligraphy}

% \begin{document} 

\begin{tikzpicture}
\tikzstyle{every node}=[font=\Large]
\begin{groupplot}[
    group style={
        group name=myplots,
        group size=1 by 12,
        vertical sep=1mm
    },
    width=15cm,
    height=2cm,
    enlargelimits=false,
    axis line style={draw=none},
    tick style={draw=none},
    xtick=\empty, ytick=\empty,
    cycle list/Dark2,
    ] 
\pgfplotsforeachungrouped \col in {1,...,12}
{
    \edef\tmp{
    \noexpand\nextgroupplot
    \noexpand\addplot+ [smooth,thick,blue!50!black] table     [col sep=comma, x=x, y index=\col] {Figures/daubechies/Daubechies_frame.csv};
    }\tmp
}
\end{groupplot}

\node (title) at ($(myplots c1r1.center)+(0,1cm)$) {Daubechies-22 Wavelet Frame};

\draw [thick, decorate,decoration={calligraphic brace,mirror,amplitude=5pt,raise=5pt}]
(myplots c1r7.north west) -- (myplots c1r12.south west) node [midway,left=10pt]{$\psi_2$};

\draw [thick, decorate,decoration={calligraphic brace,mirror,amplitude=5pt,raise=5pt}]
(myplots c1r4.north west) -- (myplots c1r6.south west) node [midway,left=10pt]{$\psi_1$};

\draw [thick, decorate,decoration={calligraphic brace,mirror,amplitude=5pt,raise=5pt}]
(myplots c1r1.north west) -- (myplots c1r3.south west) node [midway,left=10pt]{$\phi_0$};

\begin{axis}[
    title={Scaled-WaLRUS $A$ Matrix},
    name=axis2, 
    anchor=west, 
    at={(myplots c1r6.east)},  
    xshift=1cm,
    height=8cm,
    width=8cm,
    point meta min=-60,
    point meta max=60,
    enlargelimits=false,
    tick style={draw=none},
    xtick=\empty, ytick=\empty,
    colormap/PuOr,
    y dir=reverse,
    mesh/ordering=y varies,
    axis on top,
]   
    \addplot [matrix plot*, point meta=explicit] table [meta index=2] {Figures/daubechies/AdaubS.dat};

\end{axis}

\begin{axis}[
    title={Translated-WaLRUS $A$ Matrix},
    name=axis3, 
    anchor=west, 
    at={(axis2.east)},  
    xshift=1cm,
    height=8cm,
    width=8cm,
    point meta min=-60,
    point meta max=60,
    tick style={draw=none},
    enlargelimits=false,
    xtick=\empty, ytick=\empty,
    colormap/PuOr,
    y dir=reverse,
    mesh/ordering=y varies,
    axis on top,
]     
    \addplot [matrix plot*, point meta=explicit] table [meta index=2]{Figures/daubechies/AdaubT.dat}; 
\end{axis}

\begin{axis}[name=cbar, anchor=west, 
    at={(axis3.east)},
    colorbar,     
    height=8cm,
    width=2cm,
    point meta min=-60,
    point meta max=60,
    axis line style={draw=none},
    tick style={draw=none},
    axis on top,
    xtick=\empty, ytick=\empty,
    colormap/PuOr,]
    \addplot [draw=none] coordinates {(0,0)};
\end{axis}

\end{tikzpicture}

% \end{document}
    }
    \caption{Left: Elements of a Daubechies-$22$ wavelet frame, with father wavelet $\phi$, mother wavelet $\psi$, and two scales.  Right: The scaled and translated $A$ matrices for WaLRUS with $N=21$.   }
    \label{fig:daubechies_frame}
\end{figure}

%We require a frame that is differentiable, and higher-order Daubechies wavelets such as D22 (the 22-tap wavelet) meet this criteria.
%The lowest order Daubechies wavelets (e.g. Haar wavelets \cite{haar1909} (D2) and the 4-tap Daubechies (D4)) are not differentiable and thus cannot be used with SaFARi. It is notable that since WaLRUS is a SaFARi instance, any differentiable wavelet frame of choice can be used.
% \mel{suggest moving this part of the discussion to the limitations section}
% We require a differentiable frame, making higher-order Daubechies wavelets like D22 (22-tap) suitable. 
% By contrast, lower-order variants such as Haar (D2) \cite{haar1909} and D4 are nondifferentiable and therefore incompatible with SaFARi. 
% Notably, since WaLRUS is a SaFARi instance, any differentiable wavelet frame can be used.

%While it is possible to create an orthogonal basis of Daubechies wavelets with specific shifts and scales of the mother and father wavelet function, we can also modify these shifts and scales such that we have a wavelet frame with some redundancy.
%For this work, we will implement a redundant frame, following in the of generalizing SSMs beyond orthogonal bases.
Figure~\ref{fig:daubechies_frame}, left, gives a visual representation of how we construct such a frame.
The frame consists of shifted copies of the father wavelet $\phi$ at one scale, and shifted copies of a mother wavelet $\psi$ at different scales, with overlaps that introduce redundancy. Figure.~\ref{fig:daubechies_frame}, right, shows the resulting $A$ matrices for the scaled and translated WaLRUS.

\subsection{Redundancy of the wavelet frame and size of the SSM}\label{sec:frame_redundancy}
In contrast to orthonormal bases, redundant frames allow more than one way to represent the same signal. This redundancy arises from the non-trivial null space of the associated frame operator, meaning that multiple coefficient vectors can yield the same reconstructed function. 
Although the representation is not unique, it is still perfectly valid, and this flexibility offers several key advantages in signal processing. 
In particular, redundancy can improve robustness to noise, enable better sparsity for certain signal classes, and enhance numerical stability in inverse problems \citep{christensen2003introduction, grochenig2001foundations, elad2006image}.

We distinguish between the total number of frame elements $N_{\rm{full}}$ and the effective dimensionality $N_{\rm{eff}}$ of the subspace where the meaningful representations reside. 
In other words, while the frame may consist of $N_{\rm{full}}$ vectors, the actual information content lies in a lower-dimensional subspace of size $N_{\rm{eff}}$. 
This effective dimensionality can be quantified by analyzing the singular-value spectrum of the frame operator \citep{christensen2003introduction, mallat_tour}.

For the WaLRUS SSMs described in this work, we first derive $A_{N_{\rm{full}}}$ using all elements of the redundant frame. 
We then diagonalize $A$ and reduce it to a size of $N_{\rm{eff}}$. 
This ensures that different frame choices, whether orthonormal or redundant, can be fairly and meaningfully compared in terms of computational cost, memory usage, and approximation accuracy.
%SaFARi's framework remains efficient and scalable even when working with redundant frames.

%\subsection{Computational efficiency and runtime complexity} \label{sec:computational_efficiency}
%This section analyzes the computational efficiency of WaLRUS compared to HiPPO-based SSMs, emphasizing its scalability in real-world applications. Additionally, we examine representation error across SSMs by visualizing their convolution kernels.

\subsection{Computational complexity of WaLRUS }

For a sequence of length $L$, scaled-SaFARi has $O(N^3 L)$ complexity due to solving an $N$-dimensional linear system at each step, while translated-SaFARi can reuse matrix inverses, and thus has $O(N^2 L)$ complexity, assuming no diagonalization \cite{babaei2025safari}.
When the state matrix $A$ is diagonalizable, the complexity reduces to $O(N L)$ and can further accelerate to $O(L)$ with parallel processing on independent scalar SSMs. 

%We empirically observe that both scaled and translated WaLRUS are stably diagonalizable, similar to Fourier-based SSMs. 
We observe that both scaled and translated WaLRUS are stably diagonalizable.
Legendre-based SSMs, on the other hand, are not stably diagonalizable \cite{gu2020hippo}. 
%When the state matrix $A$ is diagonalizable, the complexity reduces to $O(N L)$ and can further accelerate to $O(L)$ with parallel processing on independent scalar SSMs. 
Although \cite{gu2020hippo} proposed a fast sequential HiPPO-LegS update to achieve $O(N L)$ complexity, \cite{babaei2025safari} showed that it cannot be parallelized to $O(L)$. Moreover, no efficient sequential update exists for HiPPO-LegT, leaving Legendre-based SSMs at a disadvantage during inference when sequential updates are needed.

As sequence length increases, step-wise updates become a bottleneck, especially during training when the entire sequence is available upfront. 
This can be mitigated by using convolution kernels instead of sequential updates. Precomputing the convolution kernel and applying it via convolution accelerates computation, leveraging GPU-based parallelism to achieve $O(\log L)$ run-time complexity for diagonalizable SSMs. 
This optimization is feasible for both WaLRUS and Fourier-based SSMs. 
Although Legendre-based SSMs can attain similar asymptotic complexity through structured algorithms \cite{gu2022s4, gu2022train}, their nondiagonal nature prevents decoupling into $N$ independent SSMs.
%, introducing additional computational and memory overhead.

\subsection{Representation errors in the translated WaLRUS}\label{sec:translated_SSM_error}

\begin{figure}[t!]
    \centering
    \resizebox{\textwidth}{!}{%
        \input{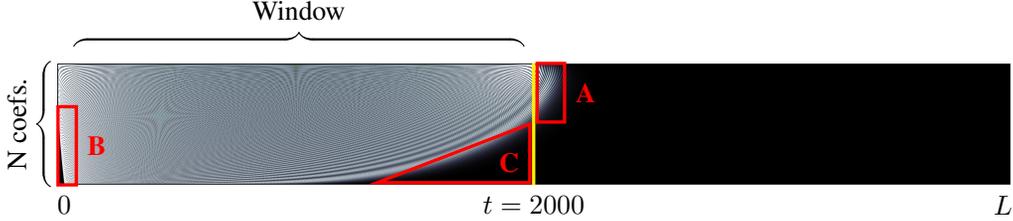}
    }
    \caption{The kernel generated by HiPPO-LegT with window size $W=2000$ and representation size $N=500$. 
    Three key non-ideal aspects of the kernel are noticeable. 
    \textbf{A)} poor localization due to substantial non-zero values outside $W$, 
    \textbf{B)} coefficient loss from at bottom left of the kernel, and   
    \textbf{C)} coefficient loss at the bottom right of the kernel for $t \in (1500,2000)$.  }\label{fig:legt-kernel}
\end{figure}

Truncated representations in SSMs inevitably introduce errors, as discarding higher-order components limits reconstruction fidelity \cite{babaei2025safari}.
%While SaFARi analyzed these errors for scaled SSMs, it did not provide error bounds for the translated variants, leaving their approximation accuracy unquantified. 
SaFARi only investigated these errors for scaled SSMs, leaving their approximation accuraty unquantified.
Visualizing the convolution kernels generated by different SSMs offers some insight into the varying performance of different SSMs on the function approximation task.
An ``ideal'' kernel would include a faithful representation for each element of the basis or frame from $T=0$ to $T=W$, where $W$ is the window width, and it would contain no non-zero elements between $W$ and $L$.
However, certain bases generate kernels with warping issues, as illustrated in Fig.~\ref{fig:legt-kernel}. 

The HiPPO-LegT kernel in Fig.~\ref{fig:legt-kernel} has substantial non-zero values outside the sliding window $W=2000$, (see area A), indicating that LegT struggles to effectively ``forget'' historical input values. 
Thus contributions from input signals outside the sliding window appear as representation errors.
Additionally, there is a loss of coefficients due to warping within the desired translating window (see areas B and C of Fig.~\ref{fig:legt-kernel}). 
For higher degrees of Legendre polynomials, the kernel exhibits an all-zero region at the beginning and end of the sliding window. 
This implies that high-frequency information in the input is not captured at the start or end of the sliding window, and the extent of this dead zone increases with higher frequencies.

\begin{figure}[t!]
    \centering
    \resizebox{\textwidth}{!}{%
        \input{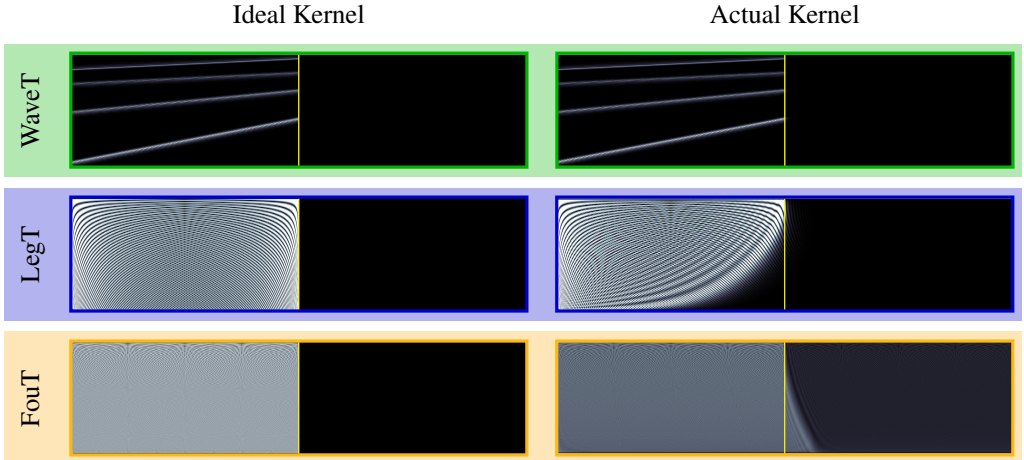}
    }  
    \caption{\textbf{Left:} The ideal kernels, which yield zero representation error, are shown for Translated-WaLRUS (using the D22 wavelet), HiPPO-LegT, and HiPPO-FouT. \textbf{Right:} The corresponding kernels generated by the translated models are presented for comparison.
    WaveT has superior localization within the window of interest compared to HiPPO-LegT and HiPPO-FouT.  }  \label{fig:translated-kernel}
    \vspace{-0.2cm}
\end{figure}

A visual inspection of Fig.~\ref{fig:translated-kernel} reveals that the translated-WaLRUS kernel closely matches the idealized version, whereas both FouT and LegT exhibit significant errors in their computed kernels. 
This is evident even for low-frequency filters, where contributions from input signals outside the sliding window contaminate the representation.
We emphasize that the issues observed with LegT and FouT arise from inherent limitations of the underlying SSMs themselves and are not due to the choice of input signal classes.

\vspace{-0.2cm}
\section{Experiments}\label{sec:experiments}
\vspace{-0.2cm}
The following section deploys the WaLRUS SSM on synthetic and real signals for the task of function approximation, comparing its performance with extant models in the literature. 
We will evaluate performance in MSE as well as their ability to track important signal features like singularities, and show that using WaLRUS can have an edge over the state-of-the-art polynomial-based SSMs. 

To benchmark WaLRUS against state-of-the-art SSMs, we implement two variants: \textit{Scaled-WaLRUS} and \textit{Translated-WaLRUS}, which we will call WaveS and WaveT respectively, following HiPPO's convention. These models are compared against the top-performing HiPPO-based SSMs. Further details on the wavelet frames used in each experiment 
are provided in Appendix~\ref{appendix:ssm_details}.

\newpage
We conduct experiments on the following datasets:

\textbf{M4 Forecasting Competition \cite{MAKRIDAKIS202054}:} A diverse collection of univariate time series with varying sampling frequencies taken from domains such as demographic, finance, industry, macro, micro, etc.

\textbf{Speech Commands \cite{SCdataset}:} A dataset of one-second audio clips featuring spoken English words from a small vocabulary, designed for benchmarking lightweight audio recognition models.

\textbf{Wavelet Benchmark Collection \cite{donoho1994ideal}:} A synthetic benchmark featuring signals with distinct singularity structures, such as Bumps, Blocks, Spikes, and Piecewise Polynomials. We generate randomized examples from each class, with further details and visualizations provided in Appendix~\ref{section:synth_examples}.

\vspace{-0.2cm}
\subsection{Comparisons among frames}
\vspace{-0.2cm}
We note that no frame is universally optimal for all input classes, as different classes of input signals exhibit varying decay rates in representation error. 
However, due to the superior localization and near-optimal error decay rate of wavelet frames, 
wavelet-based SSMs consistently show an advantage over Legendre and Fourier-based SSMs across a range of real-world and synthetic signals. 
These experiments position WaLRUS as a powerful and adaptable approach for scalable, high-fidelity signal representation.

\subsubsection{Experimental setup}
\vspace{-0.2cm}
The performance of SSMs in online function approximation can be evaluated several ways. 
One metric is the mean squared error (MSE) of the reconstructed signal compared to the original.
In the following sections, we compare the overall MSE for SSMs with a scaled measure, and the running MSE for SSMs with a translated measure.  

Additionally, in some applications, the ability to capture \textit{specific features} of a signal may be of greater interest than the overall MSE. 
As an extreme case, consider a signal that is nearly always zero, but contains a few isolated spikes.  
If our estimated signal is all zero, then the MSE will be small, but all of the information of interest has been lost.  
%This is particularly critical when the input signal contains singularities, 
% Legendre and Fourier-based SSMs tend to smooth out the input signal, discarding essential details, making them ill-equipped to handle cases where the input signal contains singularities. 
% Wavelets, by contrast, have strong responses at locations of abrupt change in a signal, making them useful for detecting signal spikes, and thus it is useful to compare the performance of SSMs built on wavelet versus polynomial frames.

In all the experiments, we use equal SSM sizes $N_{\rm{eff}}$, as described in Sec.~\ref{sec:frame_redundancy}.

\subsubsection{Function approximation with the scaled measure}

In this experiment, we construct Scaled-WaLRUS, HiPPO-LegS, and HiPPO-FouS with equal effective sizes (see Appendix~\ref{appendix:ssm_details}). Frame sizes are empirically selected to balance computational cost and approximation error across datasets.
%The average MSE for different random instances of multiple datasets is plotted in Fig.~\ref{fig:Scaled-MSE}. 

\begin{figure}[t!]
    %\centering
    \begin{minipage}[c]{0.55\textwidth} % Slightly increased width
        %\centering
        \resizebox{\textwidth}{!}{%
            \begin{tikzpicture}
\pgfplotstableread[col sep=comma,]{Figures/mse-compare/scaled_exps.csv}\datatable

\begin{axis}[
    ybar,
    error bars/y dir=both, % turn on error bars
    error bars/y explicit,  % say that error value is given explicitly
    xtick=data,
    x=1.5cm,
    xticklabels from table={\datatable}{DataType},
    ylabel={MSE},
    enlarge x limits=true,
    ]

    % --- LegS Bars ---
    \addplot [
        legColor, fill=legColor!40,
        error bars/.cd,
        y explicit,
        y dir=both,
        y explicit,
        error mark options={rotate=90, black, line width=0.8pt}
    ] table [
        x expr=\coordindex,
        y={Leg_med},
        y error minus expr={\thisrow{Leg_med} - \thisrow{Leg_low}},
        y error plus expr={\thisrow{Leg_high} - \thisrow{Leg_med}}
    ] {\datatable};
    \addlegendentry{LegS}

    % --- FouS Bars ---
    \addplot [
        fouColor, fill=fouColor!40,
        error bars/.cd,
        y explicit,
        y dir=both,
        y explicit,
        error mark options={rotate=90, black, line width=0.8pt}
    ] table [
        x expr=\coordindex,
        y={Fou_med},
        y error minus expr={\thisrow{Fou_med} - \thisrow{Fou_low}},
        y error plus expr={\thisrow{Fou_high} - \thisrow{Fou_med}}
    ] {\datatable};
    \addlegendentry{FouS}

    % --- WaveS Bars ---
    \addplot [
        daubColor, fill=daubColor!40,
        error bars/.cd,
        y explicit,
        y dir=both,
        y explicit,
        error mark options={rotate=90, black, line width=0.8pt}
    ] table [
        x expr=\coordindex,
        y={Daub_med},
        y error minus expr={\thisrow{Daub_med} - \thisrow{Daub_low}},
        y error plus expr={\thisrow{Daub_high} - \thisrow{Daub_med}}
    ] {\datatable};
    \addlegendentry{WaveS}

\end{axis}
\end{tikzpicture}
        }
        \caption{Comparing reconstruction MSE between WaveS, LegS, and FouS. Error bars represent the first and third quantile of MSE.
        WaveS produces the lowest MSE in each dataset.
        }
        \label{fig:Scaled-MSE}
    \end{minipage}%
    \hspace{0.5cm} % Pushed to the left
    \begin{minipage}{0.38\textwidth} % Slightly increased width
        \centering
        \captionof{table}{Percent of tests where each basis had the lowest overall MSE.}
        \resizebox{\textwidth}{!}{
        \begin{tabular}{c c c c} \toprule
            Dataset & LegS & FouS & WaveS \\ \midrule
            M4 & 0\% & 0.47\% & \textbf{99.53\%} \\
            Speech & 4.25\% & 0\% & \textbf{95.75\%} \\ 
            Blocks & 0\% & 0\% & \textbf{100\%} \\  
            Bumps & 0\% & 0\% & \textbf{100\%} \\ 
            Piecepoly & 1.00\% & 0\% & \textbf{99.00\%} \\ 
            Spikes & 0\% & 0\% & \textbf{100\%} \\ \bottomrule
        \end{tabular} }
        
        \label{table:percentwin}
    \end{minipage}
    \vspace{-0.4cm}
\end{figure}

Fig.~\ref{fig:Scaled-MSE} shows the average MSE across random instances of multiple datasets.
%, where WaLRUS consistently outperforms LegS and FouS. 
% WaLRUS consistently provides the lowest MSE.
% For further insight, the percentage of instances where each SSM had the best performance is also provided in Table~\ref{table:percentwin}.
Not only is the average MSE lowest for WaLRUS for all datasets, but even where there is high variance in the MSE, all methods tend to keep the same \textit{relative} performance.
That is, the overlap in the error bars in Fig.~\ref{fig:Scaled-MSE} does not imply that the methods are indistinguishable; rather, for a given instance of a dataset, the MSE across all three SSM types tends to shift together, maintaining the MSE ordering WaveS < LegS < FouS. 
To highlight this result, the percentage of instances where each SSM had the best performance is also provided in Table~\ref{table:percentwin}.

The representative power of WaLRUS is attributed to its ability to minimize truncation and mixing errors by selecting frames that capture signal characteristics with higher fidelity.  
See \cite{babaei2025safari} for further details.
% As discussed in \cite{babaei2025safari}, WaLRUS's performance is attributed to its ability to minimize truncation and mixing errors by selecting frames that capture signal characteristics with higher fidelity.

%Following discussions in \cite{babaei2025safari}, both truncation and mixing errors are mitigated by ensuring that the chosen frame captures the input signal with a high fidelity. 
%scaled-WaLRUS demonstrates a consistent advantage over LegS and FouS on our chosen datasets, pointing to the usefulness of being able to select the most appropriate basis or frame for one's signal in an SSM.

\subsubsection{Peak detection with the scaled measure}\label{sec:peak_scaled_exp}

In this experiment, we aim to detect the locations of random spikes in input sequences using Scaled-WaLRUS, FouS, and LegS, all constructed with an equal sizes. We generate random spike sequences, add Gaussian noise (SNR = 0.001), and compute their representations with Daubechies wavelets, Legendre polynomials, and Fourier series. The reconstructed signals are transformed into wavelet coefficients, and spike locations are identified following the method in \cite{mallat1992singularity}.

%In this experiment, we aim to identify the locations of random spikes in the input sequence using the generated representations. As before, we construct Scaled-WaLRUS, FouS, and LegS, each with an equal size of $N=64$. We generate instances of random spike sequences and add a small amount of Gaussian noise (SNR = 0.001). 

%For each sequence, we first compute the representation using the different frames and bases: Daubechies wavelets, Legendre polynomials, and Fourier series, and reconstruct the input sequence for each. We then transform the reconstructed signal into wavelet coefficients, and follow the method described in \cite{mallat1992singularity} to detect the spike locations.While wavelets are used to identify singularities, it is not necessarily true that a wavelet-based SSM should outperform other SSMs in this task. 

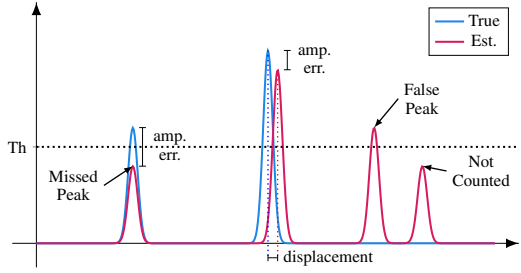
\begin{figure}[t!]
    \centering
    \resizebox{0.5\textwidth}{!}{%
        % \documentclass{standalone}
% \usepackage{tikz,pgfplots}
% %\usetikzlibrary{backgrounds,arrows.meta,hobby}
% %\usetikzlibrary{shapes.multipart}
% \usepgfplotslibrary{groupplots}
% \usepackage{amssymb}
% \usepackage{pifont}
% \pgfplotsset{compat=1.18}
% \tikzset{>=latex}
% \usetikzlibrary{arrows, decorations.markings}

% \begin{document}
%\newcommand{\supsub}[3]{{#1}\mbox{$^{#2}_{#3}$}}
%\newcommand\gauss[2]{1/({#2}*sqrt(2*pi))*exp(-((x-{#1})^2)/(2*{#2}^2))} 

\begin{tikzpicture}[every text node part/.style={align=center}]
\newcommand\gauss[2]{(1/(#2*(sqrt(2*pi))))*exp(-0.5*(((x-#1)/#2)^2))}
% \pgfplotsset{
%     no marks,axis lines*=middle,
%     inner axis line style={-{Latex[length=4mm]}}
% }
% These are colors chosen specifically for their ability to be read by colorblind
% https://davidmathlogic.com/colorblind
\definecolor{cMagenta}{HTML}{D81B60}
\definecolor{cGold}{HTML}{E0A800}
\definecolor{cLightBlue}{HTML}{1E88E5}
\definecolor{cForest}{HTML}{004D40}

\def\pwid{0.15}

\makeatletter
\tikzset{
    nomorepostaction/.code=\makeatletter\let\tikz@postactions\pgfutil@empty, % From https://tex.stackexchange.com/questions/3184/applying-a-postaction-to-every-path-in-tikz/5354#5354
    my axis/.style={
        postaction={
            decoration={
                markings,
                mark=at position 1 with {
                    \arrow[ultra thick]{latex}
                }
            },
            decorate,
            nomorepostaction
        },
        thin,
        -, % switch off other arrow tips
        every path/.append style=my axis % this is necessary so it works both with "axis lines=left" and "axis lines=middle"
    }
}
\makeatother

\begin{groupplot}[group style=
    {group size=2 by 1,     
    },
      domain=0:5,samples=500,
      xmin=-0.5, xmax=10,
      ymin=-0.5, ymax=5,
      axis lines=center,
      anchor=origin,x=1cm,y=1cm, % coincide with TikZ coordinates
      xtick=\empty,
      ytick={2},
      yticklabels={Th},
      no markers,
      axis line style={my axis},
    ]

    \nextgroupplot%[title={Amp. Error, Displacement, False/Missed Peaks}]
    \addplot [color=cLightBlue, very thick] plot [domain=0:9.5] 
    {.6*\gauss{2}{0.1}+\gauss{4.8}{0.1}};
    \addplot [color=cMagenta, very thick] plot [domain=0:9.5] 
    {.4*\gauss{2}{0.1}+0.9*\gauss{5}{0.1}+0.6*\gauss{7}{0.1}+0.4*\gauss{8}{0.1}};
    \draw [dotted,very thick] (0,2) -- (10,2);
    \node at (7.5,3) [anchor=west] {False \\ Peak};
    \draw[thick,->] (7.5,3) -- (7,2.4);
    \node at (1.4,1.2) [anchor=east] {Missed \\ Peak};
    \draw[thick,->] (1.4,1.2) -- (2,1.6);
    \node at (8.5,1.5) [anchor=west] {Not \\ Counted};
    \draw[thick,->] (8.5,1.5) -- (8,1.6);
    \addplot [scatter, only marks,
        error bars/.cd,
        y dir=both,y explicit,
    ] coordinates {
        (2.2,2.0) +- (0,0.4)
        (5.2,3.8) +- (0,0.2)
    }; 
    \node at (2.3,2.0) [anchor=west] {amp. \\ err.};
    \node at (5.3,3.8) [anchor=west] {amp. \\ err.};
    \draw[thick,dotted,blue] (4.8,4) -- (4.8,-0.3);
    \draw[thick,dotted,red] (5,3.6) -- (5,-0.3);
    \addplot [scatter, only marks,
        error bars/.cd,
        x dir=both, x explicit,
    ] coordinates {
        (4.9, -0.3) +- (.1,0)
    }; 
    \node at (5,-0.3) [anchor=west] {displacement};
    \legend{True, Est.}

    % \nextgroupplot [title={Instance-Wise Win Percentage},
    % xtick={1,3,5,7,9},
    % xticklabels={
    % \cmark A \\ \xmark B \\ \cmark C, 
    % \cmark A \\ \xmark B \\ \cmark C, 
    % \cmark A \\ \cmark B \\ \cmark C,
    % \cmark A \\ \xmark B \\ \xmark C,
    % A=100\% \\ B=25\% \\ C=75\%}
    % ]
    % \draw [dotted,very thick] (0,2) -- (10,2);
    % \addplot [color=cLightBlue, very thick] plot [domain=0:9.5] {1.2*\gauss{1}{\pwid}+1*\gauss{3}{\pwid}+1.4\gauss{5}{\pwid}+1.3\gauss{7}{\pwid}};
    % \addplot [color=cMagenta, very thick] plot [domain=0:9.5] {1.1*\gauss{1}{\pwid}+0.9*\gauss{3}{\pwid}+1.2*\gauss{5}{\pwid}+1.1*\gauss{7}{\pwid}};
    % \addplot [color=cGold, very thick] plot [domain=0:9.5] {0.7*\gauss{1}{\pwid}+0.55*\gauss{3}{\pwid}+0.95*\gauss{5}{\pwid}+0.6*\gauss{7}{\pwid}};
    % \addplot [color=cForest, very thick] plot [domain=0:9.5] {1.0*\gauss{1}{\pwid}+0.85*\gauss{3}{\pwid}+1.1*\gauss{5}{\pwid}+0.7*\gauss{7}{\pwid}};
    % \legend{True, Est. A, Est. B, Est. C}
    
\end{groupplot}

\end{tikzpicture}

% \end{document}
    }
\caption{Illustration of the metrics to evaluate performance of SSMs on different datasets in Table~\ref{table:scaled_translated_comparison}. }
\label{fig:spike_metrics}
\end{figure}

\begin{table}[t!]
\centering
\caption{Performance comparison of WalRUS-Wavelets, HiPPO-Legendre, and HiPPO-Fourier for peak detection with the translated measure. 
WalRUS shows a significant advantage in successfully remembering singularities over HiPPO SSMs.}
\vspace{6pt}
\resizebox{\textwidth}{!}{ % Resize to fit within text width
\begin{tabular}{l | c | c c c | c c c} 
\toprule
Measure & Dataset & \multicolumn{3}{c|}{Spikes} & \multicolumn{3}{c}{Bumps} \\ 
 & Basis/Frame & Legendre & Fourier & Wavelets & Legendre & Fourier & Wavelets \\ \midrule
\multirow{5}{*}{Scaled} 
& Peaks missed & 2.5\% & 0.62\% & \textbf{0\%} & 0.29\% & 0.30\% & \textbf{0\%} \\
& False peaks & 1.6\% & 1.6\% & \textbf{0.01\%} & 0.3\% & 1.9\% & \textbf{0\%} \\
& Instance-wise wins & 76\% & 92.9\% & \textbf{100\%} & 97.1\% & 96.9\% & \textbf{100\%} \\
& Relative amplitude error & 16.2\% & 11.8\% & \textbf{5.5\%} & 12.4\% & 16.2\% & \textbf{6.5\%} \\ 
& Average displacement & 18.8 & 32.0 & \textbf{10.0} & 12.7 & 33.7 & \textbf{7.1} \\ \midrule

\multirow{5}{*}{Translated}
& Peaks missed & 6.4\% & 13.0\% & \textbf{0.27\%} & 1.12\% & 29.76\% & \textbf{0.08\%} \\
& False peaks & 1.1\% & \textbf{0.05\%} & 0.22\% & 0.43\% & 0.28\% & \textbf{0.20\%} \\
& Instance-wise wins & 36.9\% & 13.65\% & \textbf{99.95\%} & 85.1\% & 0.2\% & \textbf{100\%} \\
& Relative amplitude error & 19.6\% & 28.4\% & \textbf{3.5\%} & 6.9\% & 28.4\% & \textbf{2.5\%} \\ 
& Average displacement & 6.0 & 5.4 & \textbf{4.3} & 5.5 & 5.8 & \textbf{4.8} \\
\bottomrule
\end{tabular}}

\label{table:scaled_translated_comparison}
\vspace{-0.4cm}
\end{table}

To evaluate performance, we compare the relative amplitude and displacement of detected spikes with their ground truth (see Fig.\ref{fig:spike_metrics}). This process is repeated for 1000 random sequences, each containing 10 spikes. Table \ref{table:scaled_translated_comparison} summarizes the average number of undetected spikes for each SSM and the instance-wise win percentage, representing  the number of instances where each SSM had fewer or equal misses peaks than the other SSMs. Note that these percentages do not sum to 100, as some instances result in identical spike detection across all models.

%To evaluate the performance of each SSM, after identifying the spike locations, we compare the relative amplitude of the reconstructed spikes to their ground truth values, as well as the relative displacement of the reconstructed spikes (see Fig.~\ref{fig:spike_metrics}).

%The procedure is repeated for $1000$ random sequences, each containing 10 spikes. The average number of spikes per sequence that remain undetected in the representation produced by each SSM is shown in Table~\ref{table:scaled_translated_comparison}.  Additionally, for each instance, the number of undetected spikes is compared across the different SSMs. For each SSM, we calculate the number of instances where it has the fewest missed spikes. This count is expressed as the instance-wise win percentage in Table~\ref{table:scaled_translated_comparison}. We note that these percentages do not sum to 100, as there are instances where all SSMs detect the same number of spikes.
%Fig.~\ref{fig:teaser} illustrates an example of a sequence where LegS and FouS fail to detect spikes that are closely spaced, highlighting the superior time resolution of EaLRUS.

As shown in Table~\ref{table:scaled_translated_comparison}, WaveS misses significantly fewer spikes than FouS and LegS, with lower displacement errors and reduced amplitude loss. Figure~\ref{fig:teaser} illustrates an example where WaLRUS successfully captures closely spaced spikes that are missed by LegS and FouS, demonstrating its superior time resolution.

% Overall, WaveS outperforms FouS and LegS on spike-detection tasks, achieving fewer misses, more accurate placement, and better amplitude fidelity.
%As shown in Table~\ref{table:scaled_translated_comparison}, the representation generated by scaled-WaLRUS misses significantly fewer peaks compared to FouS and LegS. Additionally, WaLRUS demonstrates reduced peak displacement and less amplitude loss in capturing peaks. The reader is referred to  Fig.~\ref{fig:teaser}  illustrating an example of a sequence where LegS and FouS fail to detect spikes that are closely spaced. In summary, scaled-WaLRUS outperforms FouS and LegS on sequences with singularities.  It not only achieves lower MSE in linear reconstruction, but also captures peaks with fewer misses, more accurate placement, and better amplitude fidelity.

\subsubsection{Function approximation with the translated measure}
\begin{figure}[t!]
    \centering
    \resizebox{0.9\textwidth}{!}{%
        \input{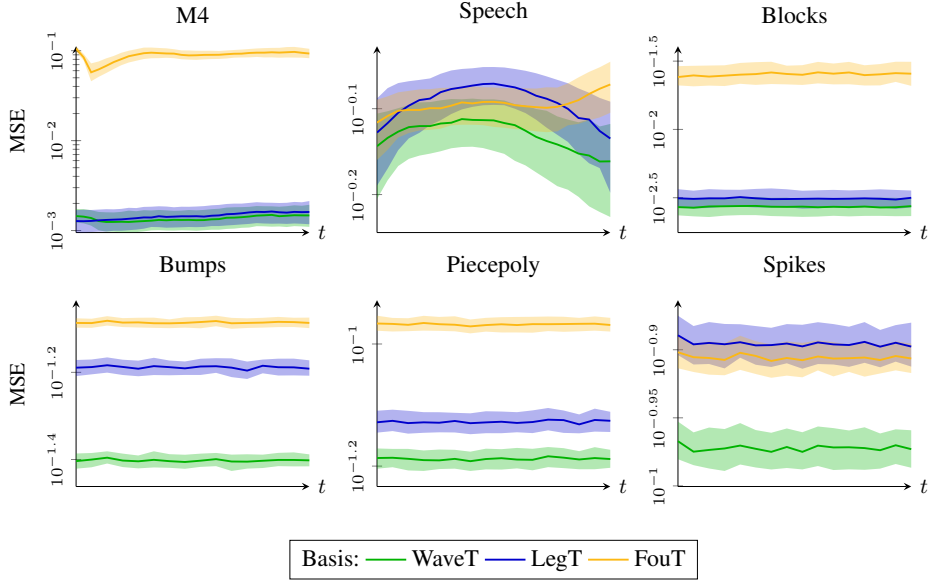}
    }
    \caption{For each dataset, the median and $(0.4,0.6)$ quantile of running reconstruction MSE across different instances is demonstrated in different colors for WaveT, LegT, and FouT. WaveT captures information in the input signals with a higher fidelity than LegT and FouT.}
    \label{fig:Translated-MSE}
\end{figure}

In this experiment, we construct WaveT, LegT, and FouT SSMs, all with equal effective sizes (see Appendix~\ref{appendix:ssm_details}).  The chosen effective sizes are smaller than those we used for the scaled measure since the translated window contains lower frequency content within each window, making it possible to reconstruct the signal with smaller frames. 
Then, for each instance of input signal, the reconstruction MSE at each time step is calculated and plotted in Fig.~\ref{fig:Translated-MSE}.

For each input signal instance, we compute the running MSE at each time step, as shown in Fig.~\ref{fig:Translated-MSE}. This plot represents how the MSE evolves over time across multiple instances, providing a comparison of running MSEs for each SSM. The results demonstrate that Translated-WaLRUS consistently achieves slightly better fidelity than LegT and significantly outperforms FouT across all datasets.

%This experiment compares the running MSE across different SSMs. To illustrate how the average MSE evolves over time for each SSM, we plot the median MSE across multiple instances at each time step.  The results indicate that translated-WaLRUS captures information in input signals across various datasets with slightly higher fidelity than LegT and significantly better fidelity than FouT.

As discussed in Section~\ref{sec:translated_SSM_error}, the reconstruction error stems from two main factors: (1) non-idealities in the translated SSM kernel, affecting its ability to retain relevant information within the window while effectively forgetting data outside it (see Fig.~\ref{fig:legt-kernel}), and (2) the extent to which these fundamental non-idealities are activated by the input signal. For example, signals with large regions of zero values are less impacted by kernel inaccuracies, as the weights outside the kernel contribute minimally to reconstruction.

%Per Section~\ref{sec:translated_SSM_error}, we note that the reconstruction error has two sources.
%The first is non-idealities present in the translated SSM kernel impacting its ability to both account for all data within the window, and effectively forget data outside of it (see Fig.~\ref{fig:legt-kernel}).
%The second factor is the extent to which the input data activates or manifests these fundamental issues, resulting in reconstruction error.
%As an example, for a signal that is mostly zero, then weights outside the kernel are likely to be multiplied by zero, and thus have little impact on the reconstruction.

WaveT achieves a modest, and in some cases negligible MSE improvement over LegT (e.g., M4 and Blocks). 
However, the kernel-based limitations highlighted in Section~\ref{sec:translated_SSM_error} may have a more pronounced effect on longer sequences or different datasets.

%Translated-WaLRUS showed a small (or in some cases, negligible, as in M4 and Blocks) MSE advantage over LegT for the datasets used in this study. However, we note that the kernel-based issues described in Section~\ref{sec:translated_SSM_error} may have a stronger impact on longer sequences or different data.

\subsubsection{Peak detection with the translated measure}
%In this experiment, we evaluate the ability of of WaLRUS, FouT, and LegT to retain information about singularities in signals. The setup follows the same methodology described in the Section~\ref{sec:peak_scaled_exp}, except with a translated SSM instead of a scaled SSM.

In this experiment, we evaluate the ability of WaveT, FouT, and LegT to retain information about singularities in signals, following the setup in Section~\ref{sec:peak_scaled_exp}, but with a translated SSM.
We generate 2,000 random sequences, each containing 20 spikes. The average number of undetected spikes for each SSM, along with instance-wise win percentages, is reported in Table~\ref{table:scaled_translated_comparison}. As in the scaled measure experiment, the percentages do not sum to 100 due to ties across SSMs.
%The procedure is repeated for 2,000 random sequences, each containing 20 spikes. The average number of undetected spikes per sequence for each SSM is summarized in Table~\ref{table:scaled_translated_comparison}, alongside the instance-wise win percentages, calculated similarly to the scaled measure experiment. Again, these percentages do not sum to 100 due to cases where all SSMs detect an equal number of spikes.
Table~\ref{table:scaled_translated_comparison} shows that WaveT consistently outperforms FouT and LegT, with fewer missed peaks, reduced displacement, and less amplitude loss.

%As indicated in Table~\ref{table:scaled_translated_comparison}, translated-WaLRUS outperforms FouT and LegT by missing significantly fewer peaks.  translated-WaLRUS also exhibits reduced peak displacement and less amplitude loss.

\section{Limitations}

In this work we have implemented only one type of wavelet (Daubechies-22), as our purpose is to introduce practical and theoretical reasons to replace polynomial SSMs with wavelet SSMs.  
Other wavelets (biorthogonal, coiflets, Morlets, etc.) could also be used, with some caveats.
First, we require a differentiable frame \cite{babaei2025safari}, so nondifferentiable wavelets like Haar wavelets or other lower-order Daubechies and Coiflets cannot be used with this method.
Second, the redundancy of the frame (and the resulting $N_{\rm{eff}}$ of the $A$ matrix) depends on the shape of the wavelet's function and the chosen shifts and scales of this function.
Other wavelet types, and other choices of shift and scale, may exhibit better or worse performance and dimensionality reduction, and this is an important question for future work.

Additionally, we emphasize that the choice of frame is application-dependent.  
If the signal is known to be smooth and periodic, a wavelet-based SSM is not likely to outperform a Fourier-based SSM, for example.  
The introduction of WaLRUS is not intended to be a one-size-fits-all model, but rather a broadly-applicable tool that combines compressive online function-approximation SSMs with the expressive power of wavelets.

% \mel{Discuss: 
% \begin{itemize}
%     \item frames must be differentiable/invertible (can't use Haar)
%     \item numerical accuracy (requires longer length of frame, unlike the closed-form orthogonal solutions)
%     \item limitations of what we actually did here? only one type of wavelet, not tested in the presence of high noise or intereference
%     \item assumptions: wavelets aren't always better than polynomials for every possible dataset, application-dependent
% \end{itemize}}

\section{Conclusions}\label{sec:conclusions}

We have demonstrated in this paper how function approximation with SSMs, initially proposed by \cite{gu2020hippo} and subsequently extended to general frames, can be improved using wavelet-based SSMs.
SSMs constructed with wavelet frames can provide higher fidelity in signal reconstruction than the state-of-the-art Legendre and Fourier-based SSMs over both scaled and translated measures.
Future work will explore alternate wavelet families, and the trade-offs in effective size, frequency space coverage, and representation capabilities of different frames.  

Moreover, since the Legendre-based HiPPO SSM forms the core of S4 and Mamba, and WaLRUS provides a drop-in replacement for HiPPO, WaLRUS could be used to initialize SSM-based machine learning models--potentially providing more efficient training.
As AI becomes ubiquitous, and the demand for computation explodes, smarter and more task-tailored ML architectures can help mitigate the strain on energy and environmental resources.

%\section{Acknowledgments}
%Special thanks to T. Mitchell Roddenberry for fruitful conversations insights over the course of this project. 
%This work was supported by NSF grants CCF-1911094, IIS-1838177, and IIS-1730574; ONR grants N00014-18-12571, N00014-20-1-2534, N00014-18-1-2047, and MURI N00014-20-1-2787; AFOSR grant FA9550-22-1-0060; DOE grant DE-SC0020345; and a Vannevar Bush Faculty Fellowship, ONR grant N00014-18-1-2047, Rice Academy of Fellows, and Rice University and Houston Methodist 2024 Seed Grant Program.

\bibliography{ref}
\bibliographystyle{unsrt}

%%%%%%%%%%%%%%%%%%%%%%%%%%%%%%%%%%%%%%%%%%%%%%%%%%%%%%%%%%%%

\newpage

\section{Appendix}
\renewcommand{\theequation}{\thesection.\arabic{equation}}
\setcounter{equation}{0}

\subsection{Additional Experimental Results}

\subsubsection{Datasets}
In this paper, we conducted our experiments on these datasets:

\textbf{M4 forecasting competition:} The M4 forecasting competition dataset \citep{MAKRIDAKIS202054} consists of 100,000 univariate time series from six domains: demographic, finance, industry, macro, micro, and other. The data covers various frequencies (hourly, daily, weekly, monthly, quarterly, yearly) and originates from sources like censuses, financial markets, industrial reports, and economic surveys. It is designed to benchmark forecasting models across diverse real-world applications, accommodating different horizons and data lengths. 
We test on 3,000 random instances.

\textbf{Speech commands:} The speech commands dataset \citep{SCdataset} is a set of 400 audio files, each containing a single spoken English word or background noise with about one second duration. 
These words are from a small set of commands, and are spoken by a variety of different speakers. 
This data set is designed to help train simple machine learning models.

\textbf{Wavelet benchmark collection:} Donoho \citep{donoho1994ideal} introduced a collection of popular wavelet benchmark signals, each designed to capture different types of singularities. 
This benchmark includes well-known signals such as  Bumps, Blocks, Spikes, and Piecewise Polynomial. 
Following this model, we synthesize random signals belonging to the classes of bumps, blocks, spikes, and piecewise polynomials. Details and examples of these signals can be found in Appendix~\ref{section:synth_examples}.

\subsubsection{Wavelet Benchmark Collection} \label{section:synth_examples}

Donoho \citep{donoho1994ideal} introduced a collection of popular wavelet benchmark signals, each designed to capture different types of singularities. 
This benchmark includes well-known signals such as Bumps, Blocks, Spikes, and Piecewise Polynomial. 

Following this model, we synthesize random signals belonging to the classes of bumps, blocks, spikes, and piecewise polynomials in our experiments to compare the fidelity of DaubS to legS and fouS, and also to compare the fidelity of DaubT to LegT and FouT.

\begin{figure}
    \includegraphics[width=.5\textwidth]{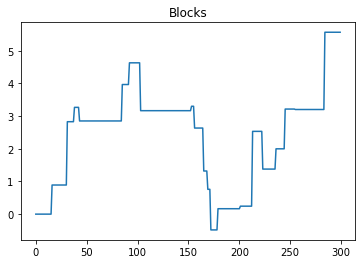}\hfill
    \includegraphics[width=.5\textwidth]{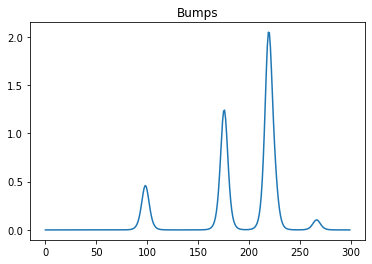}\hfill
    \\[\smallskipamount]
    \includegraphics[width=.5\textwidth]{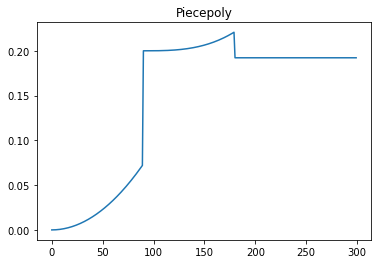}\hfill
    \includegraphics[width=.5\textwidth]{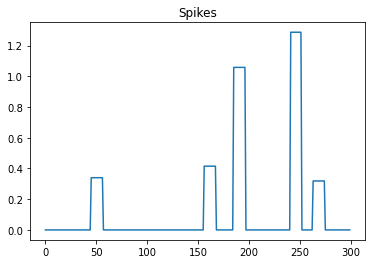}\hfill
    \caption{Instances from different types of signals in the wavelet collection benchmark that we synthesize for our experiment. \textbf{Top Left:} Blocks is a piecewise constant signal with random-hight sharp jumps placed randomly. \textbf{Top Right:} Bumps is a collection of random pulses where each pulse contains a cusp. \textbf{Bottom Left:} Piecepoly is a piecewise polynomial signal with discontinuity in the transition between different polynomial parts. \textbf{Bottom Right:} Spikes is a collection of rectangular pulses placed randomly with random positive hieght.}
    \label{fig:synth_example}
\end{figure}

Figure~\ref{fig:synth_example} demonstrates a random instance from each of of the classes of the signals that we have in our wavelet benchmark collection.

\subsubsection{Wavelet frames used for each experiment} \label{appendix:ssm_details}

Unlike HiPPO-based SSMs, which are fully characterized by their state size $N$, WaLRUS employs redundant wavelet frames that require additional parameters for identification. Once the wavelet frame is defined, the SaFARi framework constructs the unique $A, B$ matrices corresponding to that frame. The key parameters for specifying a redundant wavelet frame in WaLRUS are as follows:

\begin{itemize}
    \item \textbf{Wavelet Function:} Wavelet frames are built from a mother wavelet and a father wavelet, which capture high-frequency details and low-frequency approximations, respectively. Different families such as Daubechies, Morlet, Symlet, and Coifflet provide varied wavelet functions. For this work, we use the D22 wavelet from the Daubechies family.

    \item \textbf{L (Frame Length):} This represents the length of the wavelet frame. Increasing $L$ reduces numerical inaccuracies within the SaFARi framework.

    \item \textbf{Min Scale} and \textbf{Max Scale:} A redundant wavelet frame spans multiple dilation levels. At each level $i$, the wavelet is scaled by $2^i$, producing wavelets of length $2^i L$. Note that  we include all dilation levels within the range  Min scale and Max scale inclusively for the mother wavelet, while only the Max scale is used for the father wavelet.

    \item \textbf{Shift:} At scale $i$, $2^{-i} m$ overlapping shifts are applied to the wavelets, where $0 < m \leq 1$ is a shift constant. Setting $m = 1$ corresponds to dyadic shifts. Since our wavelet frames typically only contain a few dilation levels, the using $m=1$ often results the wavelet frame to be lossy, meaning that it does not satisfy the frame condition anymore. 

    \item \textbf{rcond:} This parameter controls the numerical stability of the pseudo-inverse calculation for the dual frame. Singular values smaller than $\rm{rcond} \times \sigma_{\rm{max}}$ are discarded during the inversion process to maintain numerical stability.
\end{itemize}

Note that all the above parameters are solely to identify the redundant wavelet frame, and that WaLRUS does not introduce any new parameters.

Table~\ref{table:wavelet_frame_details} summarizes the settings for all experiments, alongside the SSM sizes for HiPPO-Legendre and HiPPO-Fourier.

\begin{table}[t!]
\centering
\caption{ Parameters for the redundant wavelet frame used by WaLRUS in different experiments. All of the above experiment share the parameters $L=2^{19}$, and $\rm{rcond}=0.01$.  }
\vspace{6pt}
\begin{tabular}{l | c | c  c c c c } 
\toprule
Experiment &  Basis/Measure & Wavelet & scale max & scale min & shift &  $N_{\rm{eff}}$ \\ \midrule

\multirow{3}{*}{Scaled m4} 
& WaveS & D22 & 2 & -3 & 0.01 & 501\\
&  LegS & - & - & - & - &   500 \\
& FouS & - & - & - & - &  500 \\
\midrule 

\multirow{3}{*}{Scaled Speech} 
& WaveS & D22 & 0 & -5 & 0.01 & 1995\\
&  LegS & - & - & - & - &   1995 \\
& FouS & - & - & - & - &   1995 \\
\midrule

\multirow{3}{*}{Scaled synthetic} 
& WaveS & D22 & 2 & -3 & 0.01& 501\\
&  LegS & - & - & - & - & 500 \\
& FouS & - & - & - & - &   500 \\
\midrule

\multirow{3}{*}{Scaled peak detection} 
& WaveS & D22 & 3 & 0 &  0.01& 65\\
&  LegS & - & - & - & - &   65 \\
& FouS & - & - & - & - & 65 \\
\midrule

\multirow{3}{*}{Translated m4} 
& WaveT & D22 & 1 & -1 & 0.01 & 128\\
&  LegT & - & - & - & - &   128 \\
& FouT & - & - & - & - &  128 \\
\midrule 

\multirow{3}{*}{Translated Speech} 
& WaveT & D22 & 1 & -3 & 0.0025 & 500\\
&  LegT & - & - & - & - &   500 \\
& FouT & - & - & - & - &   500 \\
\midrule

\multirow{3}{*}{Translated synthetic} 
& WaveT & D22 & 1 & -1 & 0.01& 128\\
&  LegT & - & - & - & - & 128 \\
& FouT & - & - & - & - &   128 \\
\midrule

\multirow{3}{*}{Translated peak detection} 
& WaveT & D22 & 1 & 0 &  0.01& 65\\
&  LegT & - & - & - & - &   65 \\
& FouT & - & - & - & - & 65 \\
\bottomrule
\end{tabular}

\label{table:wavelet_frame_details}
\end{table}

\subsubsection{Computational resources} \label{appendix:comp_resources}

Within the scope of this paper, no networks were trained and no parameters were learned. 
This requires only CPU resources. 
As discussed in the paper, using parallel processing such as GPU speeds up the processing. But all the experiments are conducted and are replicable on CPU only. For all of our experiments, the time series we have used are scanned in the span of seconds.

Using WaLRUS to find representation has two different stages:
\begin{itemize}
    \item Pre-computing: Computing SSM $A$ matrices and diagonalizing them.
    \item Computation: Using SSM matrices to find representations of signals.
\end{itemize}

The pre-computing stage happens only once, then SSM matrices can be stored and used. 
For all our experiments except Scaled-Speech, the pre-computing stage takes less than 10 minutes. 
For the scaled-speech, the pre-computing phase can take hours. 
However, the matrices are computed only once and then stored, so this does not change the the computation at runtime.

\end{document}